\documentclass[prc,superscriptaddress,unsortedaddress,twocolumn,showpacs,preprintnumbers,amsmath,amssymb,floatfix,dvipdfmx]{revtex4}
\usepackage[dvipdfmx]{graphicx}
\usepackage{amsmath}
\usepackage{amssymb}
\usepackage{mathrsfs}
\usepackage{multirow}
\usepackage{bm}

\def\la{\mathrel{\mathpalette\fun <}}
\def\ga{\mathrel{\mathpalette\fun >}}
\def\fun#1#2{\lower3.6pt\vbox{\baselineskip0pt\lineskip.9pt
\ialign{$\mathsurround=0pt#1\hfil##\hfil$\crcr#2\crcr\sim\crcr}}}

\newcommand{\beq}{\begin{equation}}
\newcommand{\eeq}{\end{equation}}
\newcommand{\bea}{\begin{eqnarray}}
\newcommand{\eea}{\end{eqnarray}}

\newcommand{\bfi}[1]{\mbox{\boldmath $#1$}}
\newcommand{\bfis}[1]{\mbox{\boldmath ${\scriptstyle #1}$}}

\newcommand{\vK}{{\bfi K}}

\newcommand{\vs}{{\bfi s}}
\newcommand{\vt}{{\bfi t}}
\newcommand{\vL}{{\bfi L}}
\newcommand{\vell}{{\bfi \ell}}
\newcommand{\vrr}{{\bfi r}}
\newcommand{\vR}{{\bfi R}}

\newcommand{\viK}{{\bfis K}}
\newcommand{\vir}{{\bfis r}}
\newcommand{\vit}{{\bfis t}}
\newcommand{\viR}{{\bfis R}}

\begin{document}


\title{A microscopic approach to $^{3}$He scattering}

\author{Masakazu Toyokawa}
\email[]{toyokawa@phys.kyushu-u.ac.jp}
\affiliation{Department of Physics, Kyushu University, Fukuoka 812-8581, Japan}

\author{Takuma Matsumoto}
\affiliation{Department of Physics, Kyushu University, Fukuoka 812-8581, Japan}

\author{Kosho Minomo}
\affiliation{Research Center for Nuclear Physics, Osaka University, Ibaraki 567-0047, Japan}

\author{Masanobu Yahiro}
\affiliation{Department of Physics, Kyushu University, Fukuoka 812-8581, Japan}

\date{\today}

\begin{abstract}
We propose a practical folding model to describe $^{3}$He elastic scattering. 
In the model, $^{3}$He optical potentials are constructed by making 
the folding procedure twice. First the nucleon-target potential is evaluated 
by folding the Melbourne $g$-matrix with the target density 
and localizing the nonlocal folding potential 
with the Brieva--Rook method, and second 
the resulting local nucleon-target potential is folded 
with the $^{3}$He density. 
This double single-folding model well describes $^{3}$He elastic scattering 
from $^{58}$Ni and $^{208}$Pb targets in a wide incident-energy range 
from 30 MeV/nucleon to 150 MeV/nucleon with no adjustable parameter.  
Spin-orbit force effects on differential cross sections are found 
to be appreciable only at higher incident energies such as 150 MeV/nucleon. 
Three-nucleon breakup effects of $^{3}$He are investigated with 
the continuum discretized coupled-channels method and are found to be 
appreciable only at lower incident energies around 40 MeV/nucleon. 
Effects of knock-on exchange processes are also analyzed. 
\end{abstract}


\maketitle
\section{Introduction}
\label{Introduction}

Microscopic understanding of nucleon--nucleus (NA) and 
nucleus--nucleus (AA) optical potentials is a goal of nuclear physics. 
The optical potentials are not only essential quantities 
to describe the elastic scattering but also key inputs 
in calculations of the distorted-wave Born approximation (DWBA) 
and the continuum discretized coupled-channels method (CDCC)~\cite{CDCC-review1,CDCC-review2,CDCC-review3} to describe inelastic scattering, breakup and transfer reactions.

The $g$-matrix folding model is a method of deriving the 
optical potentials microscopically. Here the $g$-matrix is 
an effective nucleon-nucleon (NN) interaction 
in nuclear matter and depends on 
the density $\rho$ of nuclear matter: $g=g(\rho)$.  
In the folding model, the optical potential is obtained 
by folding the $g$-matrix 
\cite{M3Y,JLM,Brieva-Rook,Satchler-1979,Satchler,CEG,Rikus-von
Geramb,Amos,CEG07,MP,Saliem} with the target density $\rho_{\rm T}$ 
for NA scattering and with $\rho_{\rm T}$ and the projectile one 
$\rho_{\rm P}$ for AA scattering; 
see for example Refs.~\cite{DFM-standard-form,Arellano:1995,rainbow,
DFM-standard-form-2,Sum12} for the folding procedure. 
The model is now called the single-folding (SF) model for NA scattering 
and the double-folding (DF) model for AA scattering. 
For AA scattering, occasionally, the optical potential is   
obtained semi-microscopically by folding the phenomenological 
nucleon-target potential with $\rho_{\rm P}$. 
This procedure is also called the SF model.

For NA elastic scattering, the SF model based on 
the Melbourne $g$-matrix~\cite{Amos} well reproduces 
the experimental data with no adjustable parameter. In the folding procedure, 
the value of $\rho$ in $g(\rho)$ is assumed to be a value of $\rho_{\rm T}$ 
at the midpoint $\vrr_{\rm m}$ of interacting two nucleons: 
$\rho=\rho_{\rm T}(\vrr_{\rm m})$. 
This local-density approximation seems to be reasonable because of 
the success of the Melbourne $g$-matrix folding model.

The NA potential thus obtained is non-local, 
because knock-on exchange processes are taken into account in the folding 
procedure. However, it can be localized by the Brieva--Rook method 
\cite{Brieva-Rook} with good accuracy. The validity of the method is shown 
in Refs. \cite{Minomo:2009ds,Hag06}. In fact, the
local version of the folding potential agrees with 
the phenomenological NA optical potentials determined from 
the NA scattering data~\cite{Perey-Perey,Dirac1,Dirac2,Koning-Delaroche} 
particularly in the surface region important 
for the scattering~\cite{Toyokawa:2013uua}.

The multiple NN collision series 
in AA scattering~\cite{Yahiro-Glauber} is more complicated than 
in NA scattering~\cite{Watson,KMT}. 
This makes microscopic description of AA scattering more difficult. 
In fact, the DF model has a basic problem. 
In nuclear matter calculations, in principle, 
the $g$-matrix should be derived by solving 
the scattering between a nucleon in a Fermi sphere and a nucleon 
in another Fermi sphere~\cite{Izu80}, 
but in practice it is evaluated 
by solving nucleon scattering on a single Fermi sphere. 
In addition to the single-sphere approximation, 
furthermore, in the folding procedure  
the density $\rho$ of the single Fermi sphere is assumed to be identical 
with the sum of $\rho_{\rm P}$ and $\rho_{\rm T}$ at the midpoint $\vrr_{\rm m}$ 
of interacting two nucleons: 
\bea
\rho={\rho_{\rm P}(\vrr_{\rm m})}+{\rho_{\rm T}(\vrr_{\rm m})} . 
\label{FD-approx}
\eea
The prescription is called the frozen-density approximation (FDA). 
The DF model with the FDA is referred to as the DF-FDA model in this paper.

In actual DF-FDA calculations, the magnitude of the folding potential 
is usually adjusted to the experimental data. 
It is then an important subject in future to clarify how good the DF-FDA model 
is. As a successful example, the DF-FDA model based on the 
Melbourne $g$-matrix well reproduced measured total reaction cross sections 
$\sigma_{\rm R}$ for $^{12}$C scattering from stable nuclei at around 250 MeV/nucleon 
with no adjustable parameter~\cite{Min11,Min12,Sum12,Takechi-2010,Watanabe-Mg-2014,Takechi-Mg}. The DF-FDA model was then applied to measured 
$\sigma_{\rm R}$~\cite{Takechi-2010,Takechi-Mg} for neutron-rich 
Ne and Mg isotopes~\cite{Min11,Min12,Sum12,Takechi-2010,Watanabe-Mg-2014,Takechi-Mg}. This analysis leads to the result 
that $^{31}$Ne and $^{37}$Mg are 
deformed halo nuclei.

As an alternative approach to the DF-FDA model, we can consider the model 
Hamiltonian
\bea
H_{\rm eff}=K_R + \sum_{i \in {\rm P}} U_{i{\rm T}} + h_{\rm P},
\label{model-H}
\eea
where $K_R$ stands for the kinetic energy with respect to the relative
coordinate $\vR$ between a projectile (P) and a target (T) and 
$h_{\rm P}$ is the intrinsic Hamiltonian of P. 
Here $U_{i{\rm T}}$ represents the interaction between T and 
the $i$th nucleon in P. As $U_{i{\rm T}}$, the phenomenological 
nucleon-target optical potential was used so far; 
see for example Refs. \cite{CDCC-review1,CDCC-review2,CDCC-review3} 
for deuteron scattering. 
In this paper, meanwhile, $U_{i{\rm T}}$ is constructed microscopically 
by folding the Melbourne $g$-matrix with $\rho_{\rm T}$ and is localized 
by the Brieva--Rook method. The potential $U_{i{\rm T}}$ is then 
always obtainable even if 
no experimental data is available for nucleon-target scattering of interest. 
This is an advantage of the present approach from the previous one. 
Furthermore, this approach is consistent with the fact 
that a single Fermi sphere is considered in the $g$-matrix calculation 
for nuclear matter. 
In addition, one can treat projectile breakup with CDCC if necessary, 
since $U_{i{\rm T}}$ is localized. 
These are advantages of the present approach from the DF-FDA model. 
The model Hamiltonian \eqref{model-H} 
can be derived from the many-body Hamiltonian with reasonable approximations, 
as shown later in Sec. \ref{Theoretical framework}.

It is known that the model Hamiltonian \eqref{model-H} well accounts for 
deuteron scattering, 
when deuteron breakup effects are properly taken into account 
with CDCC \cite{CDCC-review1,CDCC-review2,CDCC-review3}. 
For $^{4}$He scattering, meanwhile, projectile-excitation effects are 
quite small since $^{4}$He is hardly excited. 
We can then expect that the scattering is described 
by the optical potential that is obtained by folding 
the local version of microscopic $U_{i{\rm T}}$ 
with the $^{4}$He density. 
We refer to this model as the double single-folding (DSF) model 
in this paper. 
Very recently, it was shown that the DSF model well accounts for 
differential elastic cross sections and $\sigma_{\rm R}$ for 
$^{4}$He scattering without introducing any 
adjustable parameter~\cite{Egashira:2014zda}. 
In fact, the DSF model yields better agreement 
with the data than the DF-FDA model for $^{4}$He scattering. 
If the success of the model Hamiltonian \eqref{model-H} 
for deuteron and $^{4}$He scattering is not accidental, 
the model Hamiltonian should be good also for $^{3}$He scattering. 
This is an important question to understand AA scattering systematically.

In this paper, we investigate how good the model Hamiltonian is 
for $^{3}$He scattering, and  
show that the model Hamiltonian works well and 
the DSF model based on the model Hamiltonian is a practical model 
to describe $^{3}$He scattering. 
This analysis is made in a wide incident-energy range 
of $30\mbox{--}150$~MeV/nucleon. 
We consider heavier targets such as $^{58}$Ni and $^{208}$Pb 
in order to make our discussion clear, since the $g$-matrix is evaluated 
in nuclear matter and hence the $g$-matrix folding model is 
considered to be more reliable for heavier targets.  
$^{3}$He is more fragile than $^{4}$He, but less fragile than deuteron. 
We then investigate three nucleon breakup of $^{3}$He by using CDCC. 
In the model Hamiltonian \eqref{model-H}, 
the nucleon-target potential $U_{i{\rm T}}$ 
is local. This makes CDCC calculations feasible.  
We show that the effects are appreciable only at 
lower incident energies around 40 MeV/nucleon and negligibly small 
at higher incident energies. This makes the DSF model reliable. 
In fact, the model yields better agreement with the experimental data 
on total reaction and differential cross sections than the DF-FDA model. 
We also investigate how the spin-orbit force between $^{3}$He and T 
affects $^{3}$He scattering by using the DSF model. 
Finally we analyze effects of knock-on exchange processes on 
$^{3}$He scattering. For AA scattering, the processes make 
the microscopic optical potential nonlocal, 
but the processes are approximately treated in the DSF model, 
since $U_{i{\rm T}}$ is localized.  We then investigate 
how the approximation affects $^{3}$He scattering.

In Sec. \ref{Theoretical framework}, we derive the model Hamiltonian 
\eqref{model-H} from the many-body Hamiltonian with reasonable approximations, 
using the multiple scattering theory~\cite{Watson,KMT,Yahiro-Glauber}. 
Brief explanation is made on the DSF and DF-FDA models. 
The explicit form of the spin-orbit potential between $^{3}$He and T is shown, 
and four-body CDCC is recapitulated. 
In Sec. \ref{Results}, numerical results are shown. 
Section \ref{Summary} is devoted to a summary.

\section{Model building}
\label{Theoretical framework}

We consider the scattering of P with mass number 
$A_{\rm P}$ from T with mass number $A_{\rm T}$. 
In principle, the scattering is described by 
the many-body Schr\"odinger equation
\bea
\Bigl[
K_R +h_{\rm P}+h_{\rm T}+ \sum_{i \in {\rm P}, j
\in {\rm T}} v_{ij}-E 
\Bigr] 
{\Psi}^{(+)}=0
\label{schrodinger-bare}
\eea
for the total wave function ${\Psi}^{(+)}$, 
where $v_{ij}$ is the realistic NN interaction and $h_{\rm T}$ stands for 
the internal Hamiltonian of T. 
The total energy $E$ is related to the incident energy $E_{\rm in}^{\rm cm}$ 
in the center of mass system 
as $E=E_{\rm in}^{\rm cm}+\epsilon_{0}(\rm P)+\epsilon_{0}(\rm T)$, where 
$\epsilon_{0}(\rm P)$ and $\epsilon_{0}(\rm T)$ are 
the ground-state energies of P and T, respectively. 
Following the multiple scattering theory~\cite{Watson, KMT,Yahiro-Glauber}, 
one can rewrite Eq.~\eqref{schrodinger-bare} into 
\bea
\Bigl[
K_R +h_{\rm P}+h_{\rm T}+ \frac{Y-1}{Y} \sum_{i \in {\rm P}, j \in {\rm T}}
\tau_{ij}-E
\Bigr] 
{\hat \Psi}^{(+)}=0 , 
\label{schrodinger-effective}
\eea
where $\tau_{ij}$ denotes the effective NN interaction in nuclear medium 
and the factor $Y=A_{\rm P}A_{\rm T}$ represents the number of 
the $\tau_{ij}$ working between P and T. 
For AA scattering with $Y \gg 1$, 
the factor $(Y-1)/Y$ can be approximated into 1. 
When Eq. \eqref{schrodinger-effective} is derived 
from Eq. \eqref{schrodinger-bare}, 
the antisymmetrization between nucleons in P 
and those in T are neglected. 
However, the antisymmetrization effects are well taken care of, 
if the $\tau_{ij}$ are symmetrical with respect to the exchange 
of colliding nucleons~\cite{Takeda,Picklesimer}. 
The $\tau_{ij}$ is often replaced by the $g$-matrix ($g_{ij}$) 
in many applications, since both include nuclear medium effects.

Therefore, we reach the Schr\"odinger equation
\bea
\Bigl[
K_R +h_{\rm P}+h_{\rm T}+ \sum_{i \in {\rm P}, j \in {\rm T}}
g_{ij}-E
\Bigr] 
{\hat \Psi}^{(+)}=0 . 
\label{schrodinger-effective-g}
\eea
In this work, we use the Melbourne $g$-matrix~\cite{Amos} as $g_{ij}$. 
As mentioned in Sec. \ref{Introduction}, $g(\rho)$ is evaluated 
in nuclear matter by solving 
nucleon scattering on a single Fermi sphere. 
For consistency with the nuclear-matter calculation, 
we consider the nucleon-target subsystem in the P+T system and assume 
\bea
\rho=\rho_{\rm T}(\vrr_m) 
\label{TD-approx}
\eea
as $\rho$ in $g(\rho)$.  
This procedure is referred to as the target-density approximation (TDA) 
in this paper.

In the TDA, $g({\rho_{\rm T}})$ includes 
target-excitation effects approximately, but does not include 
projectile-excitation effects. 
We can then assume that ${\hat \Psi}^{(+)}=\phi_{0}(\rm T) \psi$, where 
$\phi_{0}(\rm T)$ is the ground state of T and $\psi$ 
describes the scattering of P from T in its ground state. 
Left-multiplying 
Eq. \eqref{schrodinger-effective-g} by $\phi_{0} (\rm T)$, 
we can get the Schr\"odinger equation
\bea
 \left[ H_{\rm eff} - E_{\rm in}^{\rm cm} - \epsilon_{0}(\rm P) \right] \psi 
= 0 
\label{schrodinger for P+T0}
\eea
for $\psi$. Here the nucleon-target potential 
\bea
U_{i{\rm T}}=\langle \phi_{0}({\rm T})| \sum_{j \in {\rm T}}
g_{ij}({\rho_{\rm T}}) | \phi_{0}({\rm T}) \rangle 
\label{Pot-NTF}
\eea
is composed of the direct and knock-on exchange terms. 
The knock-on exchange process makes $U_{i{\rm T}}$ nonlocal, but 
it can be localized with the Brieva--Rook method~\cite{Brieva-Rook} 
based on the local semi-classical approximation with good accuracy 
\cite{Minomo:2009ds}. 
In this approach, projectile excitations have to be treated 
explicitly by solving Eq. \eqref{schrodinger for P+T0}.

When projectile excitations are negligible, one can assume 
$\psi=\phi_{0}({\rm P})\chi_{0}(\vR)$ with the ground state 
$\phi_{0}({\rm P})$ of P and the relative wave function 
$\chi_{0}(\vR)$. Equation \eqref{schrodinger for P+T0} is then 
reduced to the Schr\"odinger equation 
\bea
[K_R+U_{\rm DSF}(\vR)-E_{\rm in}^{\rm cm}]\chi_{0}(\vR)=0,
\label{Schrodinger-single-channel}
\eea
where the optical potential $U_{\rm DSF}$ is obtained by 
\bea
U_{\rm DSF}(\vR)=\langle \phi_{0}({\rm P})| \sum_{i \in {\rm P}}
U_{i{\rm T}} | \phi_{0}({\rm P}) \rangle .
\label{Pot-DSF}
\eea
The potential $U_{\rm DSF}$ is local, since $U_{i{\rm T}}$ is localized. 
This folding procedure is the DSF. 
The difference between the DSF and SF models comes from the difference 
of $U_{i{\rm T}}$. In the DSF model, $U_{i{\rm T}}$ is a 
microscopic potential obtained by folding the Melbourne $g$-matrix 
with $\rho_{\rm T}$, but 
$U_{i{\rm T}}$ is a phenomenological 
nucleon-nucleus potential in the SF model. 
The Coulomb potential $U_{\rm Coul}$ is added 
to $U_{\rm DSF}$ in actual calculations. 
It is reported in Ref. \cite{Egashira:2014zda} 
that the DSF model works well for $^{4}$He scattering.

Another approach to AA scattering is the DF-FDA model. 
In this approach, the $g$-matrix $g({\rho_{\rm P}}+{\rho_{\rm T}})$ 
includes both projectile- and target-excitation effects 
approximately. The optical potential of P+T scattering is then 
obtained from Eq. \eqref{schrodinger-effective-g} as 
\bea
U_{\rm DF}^{\rm FDA}(\vR)=\langle  \Phi_0
| \sum_{i \in {\rm P}, j \in {\rm T}}g_{ij}({\rho_{\rm P}}+{\rho_{\rm T}}) 
| \Phi_0 \rangle \;
\label{Pot-DF-FDA}
\eea
for $\Phi_0=\phi_{0}(\rm P)\phi_{0}(\rm T)$. 
The potential is also composed of the direct and knock-on exchange terms. 
The nonlocality coming from the knock-on exchange process can be localized 
by the Brieva--Rook method with good accuracy \cite{Hag06}. 
In actual calculations, $U_{\rm Coul}$ is added 
to the localized $U_{\rm DF}^{\rm FDA}$. This is the DF-FDA model.

Comparing Eq. \eqref{Pot-DSF} with Eq. \eqref{Pot-DF-FDA}, one can see that 
the difference between the DSF and DF-FDA potentials mainly comes from 
that between the TDA and the FDA; see Appendix 
\ref{Explicit forms of DF-FDA, DF-TDA and DSF potentials} 
for the detail. 
Since $\rho_{\rm P}+\rho_{\rm T} > \rho_{\rm T}$, 
nuclear-medium effects are smaller in the DSF potential than in the DF-FDA 
potential. This makes the DSF potential more attractive 
and more absorptive than the DF-FDA potential.

As for the central part of DSF and DF potentials, 
we summarize the explicit forms in Appendix 
\ref{Explicit forms of DF-FDA, DF-TDA and DSF potentials}. 
In general, the $^{3}$He optical potential has the spin-orbit part 
in addition to the central part. General derivation of 
the spin-orbit part for AA scattering was shown 
in Ref. \cite{3He:Petrovich}. 
For $^{3}$He scattering, the spin-orbit part was calculated 
in Ref. \cite{3He:Cook, 3He:Sakuragi} 
for the DF model and in Ref. \cite{3He:Sakuragi} for the SF model. 
Although adjustable parameters are introduced in the analyses, 
spin-orbit effects on differential cross sections are similar 
between the two models. 
We then derive 
the spin-orbit part of $^{3}$He optical potential only in the DSF model. 
In $^{3}$He, the two-proton subsystem is considered to be spin-singlet 
with good accuracy. Hence the spin-orbit force 
$U_{\rm DSF}^{\rm LS}(\vR) \vL \cdot \vs$ of $^{3}$He scattering 
at an incident energy $E_{\rm in}$ in the laboratory system 
is obtained by folding the spin-orbit force 
$U_{\rm nT}^{LS}(\vrr) \vell \cdot \vs$ of neutron scattering 
at an incident energy $E_{\rm in}/3$   
with the neutron wave function $\phi_n(\vt)$ in $^{3}$He, 
where $\vs$ is the neutron spin, 
$\vrr$ ($\vt$) is the coordinate of neutron from the center of mass 
of T ($^{3}$He) and $\vL$ ($\vell$) is the angular momentum with respect to 
$\vR$ (\vrr). Assuming that $\phi_n(\vt)$ is an s-state, one can get 
\bea
\label{eq:DSF-LS}
U_{\rm DSF}^{\rm LS}(\vR) \vL \cdot \vs &=&
\int  \phi_n(\vt)^{*} U_{\rm nT}^{LS}(\vrr) \vell \cdot \vs \phi_n(\vt)        
d \vt, \nonumber \\
&=&  \frac{A_{\rm P}+A_{\rm T}}{A_{\rm P}(1+A_{\rm T})} \frac{1}{R}\frac{dZ(R)}{dR} \vL \cdot 
\vs
\eea
with 
\bea
Z(R)&=&\int e^{i\viK \cdot \viR}
\frac{1}{K}\frac{d {\tilde U}(K)}{dK} {\tilde \rho}(K)  d \vK, \\
{\tilde U}(K)&=&\frac{1}{(2\pi)^3} 
\int e^{-i\viK \cdot \vir} U_{\rm nT}^{LS}(\vrr) d\vrr, \\
{\tilde \rho}(K)&=&\int e^{i\viK \cdot \vit}|\phi_n(\vt)|^{2} d\vt ,
\eea
where the explicit form of $U_{\rm nT}^{LS}(\vrr)$ is shown 
in Ref. \cite{CEG07}.

Finally, three-body breakup effects of $^{3}$He are taken into account 
with four-body CDCC \cite{CDCC-review3}. 
Four-body dynamics of the $N$+$N$+$N$+T system is treated in a model space
\begin{eqnarray}
 {\cal P}=\sum_{\gamma=0}^{\gamma_{m}}|\phi_{\gamma}(\rm P) \rangle \langle 
\phi_{\gamma}(\rm P)|,
 \label{eq:com-set-s5}
\end{eqnarray}
where ${\phi}_{\gamma}(\rm P)$ is the $\gamma$th eigenstate 
with an eigenenergy $\epsilon_{\gamma}(\rm P)$ 
obtained by
diagonalizing $h_{\rm P}$ by the Gaussian basis functions and 
among the eigenstates 
the $\gamma_m$th eigenstate ${\phi}_{\gamma_m}(\rm P)$ has 
the highest eigenenergy in the model space ${\cal P}$; 
note that 
P is $^{3}$He in the present case. 
This model-space approximation reduces 
Eq. \eqref{schrodinger for P+T0} to 
\bea
{\cal P} [ H_{\rm eff} - E_{\rm in}^{\rm cm} - \epsilon_{0}(\rm P) ] {\cal P} \psi = 0,
\label{schrodinger-CDCC}
\eea
where
\bea
{\cal P}\psi =
\sum_{\gamma=0}^{\gamma_{m}} \phi_{\gamma}(\rm P) \chi_{\gamma}(\vR).
\eea
This leads to the CDCC equation for $\chi_{\gamma}(\vR)$ as 
\bea
&& \left[
E_{\rm in}^{\rm cm} 
- K_{R} - \left(\epsilon_{\gamma}(\rm P)-\epsilon_{0}(\rm P)\right) 
\right] \chi_{\gamma}(\vR)
\nonumber \\
&&~= \sum_{\gamma'}^{\gamma_{m}}
   \langle \phi_{\gamma}(\rm P)|\sum_{i \in {\rm P}} U_{i{\rm T}}
| \phi_{\gamma'}(\rm P) \rangle \chi_{\gamma'}(\vR) .
\label{CC-eq-s2}
\eea
Here the folding potential $U_{i{\rm T}}$ and the coupling potentials $
\langle \phi_{\gamma}(\rm P)|\sum_{i \in {\rm P}} U_{i{\rm T}}
| \phi_{\gamma'}(\rm P) \rangle$ are calculated with direct numerical 
integration in the coordinate space. 
The CDCC equation is numerically solved 
with kinematical relativistic corrections, 
but the corrections are negligible in the present cases. 

As shown later in Sec. \ref{Results}, $^3$He breakup effects are 
appreciable only at lower incident energies around 40 MeV/nucleon and 
negligible at higher incident energies. 
Therefore, we estimate the breakup effects as simply as possible. 
For this purpose, we assume that $^3$He is a spinless
particle, and take the Minnesota force \cite{Minnesota-force} 
as the nucleon-nucleon interaction in $^{3}$He and introduce 
a three-body force to reproduce the binding energy of $^3$He. 
In CDCC calculations, $0^+$ and $2^+$ states are considered as breakup
states, since the contribution of $1^-$ breakup states is confirmed 
to be negligibly small. 
For each of $0^+$ and $2^+$, we take 35 breakup states 
below 20 MeV. The model space spanned by the ground and breakup states 
is confirmed to give the convergence of CDCC solutions for the
elastic and reaction cross sections.

\section{Results}
\label{Results}

We consider $^{3}$He scattering from $^{58}$Ni and $^{208}$Pb targets 
in a wide incident-energy range of $30 \la E_{\rm in}/A_{\rm P} \la 150$~MeV, 
where $E_{\rm in}$ is the incident energy in the laboratory system. 
First the scattering are analyzed with the DSF and DF-FDA models in which 
the spin-orbit force is not included. 
As the $^{3}$He density ($\rho_{\rm P}$), 
we consider the density calculated with the three-nucleon model mentioned 
in Sec. \ref{Theoretical framework} and 
the phenomenological density determined 
from electron scattering~\cite{C12-density}. For the latter, 
finite-size effects due to the proton charge are unfolded in the standard manner~\cite{Singhal}, and the neutron density is assumed to have the same geometry 
as the proton one. Both the densities yield almost the same differential 
cross section, so we will take the former density to compare results 
of the DSF model with those of CDCC later.

The target density $\rho_{\rm T}$ is calculated with 
the spherical Hartree-Fock (HF) method 
with the Gogny-D1S interaction~\cite{GognyD1S} 
in which the spurious center-of-mass motion is removed 
with the standard procedure~\cite{Sum12}. 
In the present calculation, pairing effects are not included, 
but it is possible to take into account the effects with the spherical Hartree-Fock-Bogoliubov (HFB) method with Gogny-D1S force. We have confirmed that 
the effects are negligible on $\sigma_{\rm R}$ and 
differential cross sections at $q \la 5$~fm$^{-1}$ 
for $^{3}$He+$^{58}$Ni scattering at $E_{\rm in}/A_{\rm P}=72$~MeV, 
where $q$ is the transfer momentum. 
We can also deal with effects of nuclear deformation and inelastic channels 
by solving the coupled-channel equation with coupling potentials among elastic and inelastic channels. The coupling potentials can be obtained 
by folding g-matrix with the transition densities calculated 
by structure models; 
see for example Refs. \cite{Sum12,CC:Khoa,CC:Furumoto}. 

The  Melbourne interaction is provided only up to 
$k_{\rm F} = 1.5$~fm$^{-1}$ ($\rho=1.37 \rho_0$), where 
$k_{\rm F}$ is the Fermi momentum and $\rho_0$ is the normal density. 
We then assume that the Melbourne interaction at $k_{\rm F}>1.5$~fm$^{-1}$ is 
the same as that at $k_{\rm F} =1.5$~fm$^{-1}$. 
This assumption does not affect any result of DSF calculations, since 
the $g$-matrix at $k_{\rm F} > 1.5$~fm$^{-1}$ yields small effects 
only on differential cross sections at larger angles in DF-FDA calculations.

\begin{figure}[htbp]
\begin{center}
 \includegraphics[width=0.40\textwidth,clip]{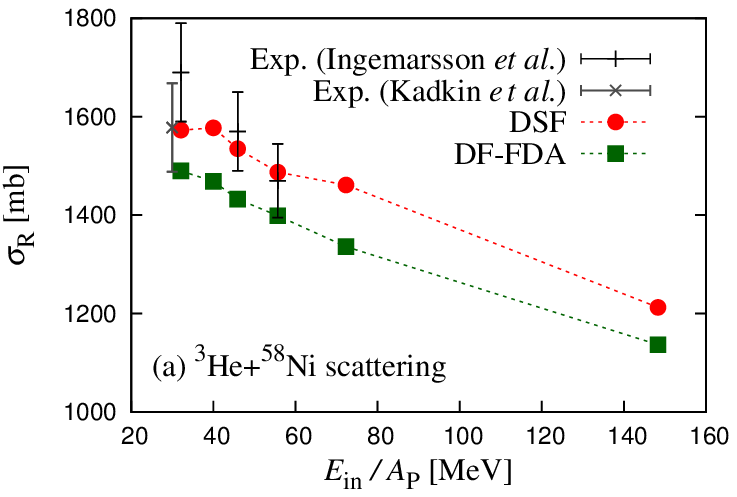}
 \includegraphics[width=0.40\textwidth,clip]{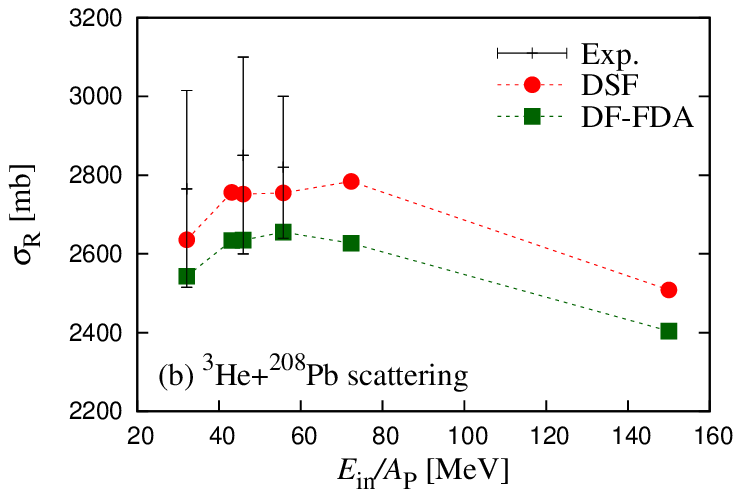}
 \caption{(Color online) Total reaction cross section  $\sigma_{\rm R}$
as a function of $E_{\rm in}/A_{\rm P}$ for (a) $^{3}$He+$^{58}$Ni scattering 
and (b) $^{3}$He+$^{208}$Pb scattering.
The circles (squares) stand for results of the DSF (DF-FDA) model. 
The spin-orbit force is not included in both the models. 
The experimental data are taken from 
\cite{E32-56:Ingemarsson:2001,E31:Kadkin:1998}. 
}
 \label{Fig-ReactionXSEC-He3Ni58-Pb208}
\end{center}
\end{figure}

\begin{figure}[htbp]
\begin{center}
 \includegraphics[width=0.40\textwidth,clip]{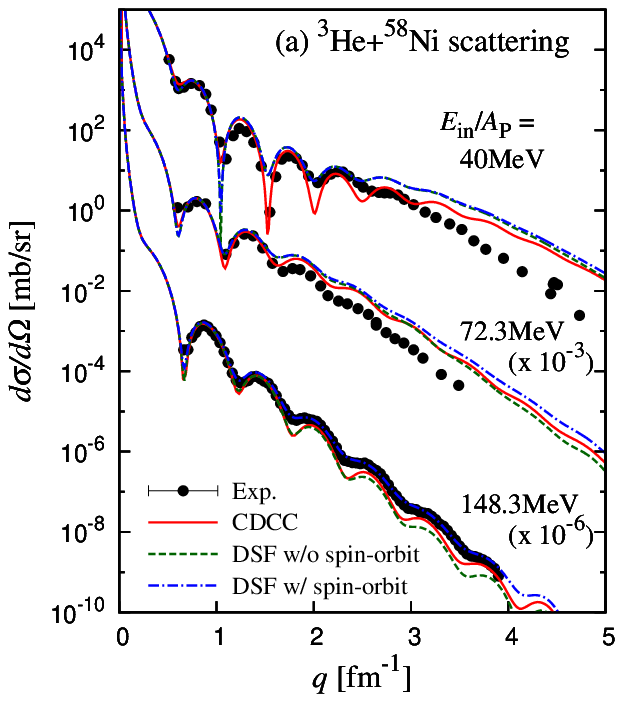}
 \includegraphics[width=0.40\textwidth,clip]{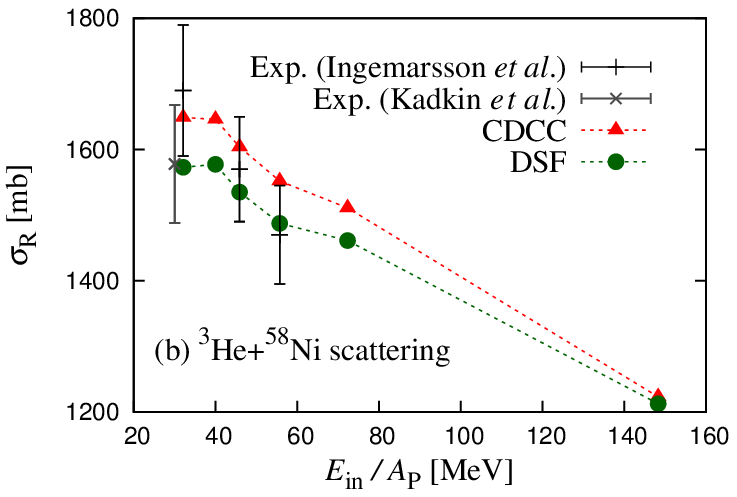}
 \caption{(Color online) 
Projectile-breakup and spin-orbit force effects 
on (a) differential cross sections 
and (b) total reaction cross sections for $^{3}$He+$^{58}$Ni scattering. 
In panel (a), the differential cross section at each $E_{\rm in}/A_{\rm P}$ is 
multiplied by the factor shown in the panel. 
The dot-dashed and dashed lines denote results of the DSF model 
with and without 
the spin-orbit force, respectively, while 
the solid lines correspond to results of CDCC in which the spin-orbit 
force is neglected. 
In panel (b), the circles stand for results of the DSF model 
without the spin-orbit force, while the triangles correspond to 
results of CDCC in which the spin-orbit force is neglected. 
Results of the DSF model with the spin-orbit force agree 
with those of the DSF model without the spin-orbit force. 
The experimental data are taken from Refs.~
\cite{E40:Hyakutake:1980,E72:Willis:1973,E148:Kamiya:2003,E32-56:Ingemarsson:2001,E31:Kadkin:1998}. 
For $E_{\rm in}/A_{\rm P}=72.3$ MeV in panel (a), 
the CDCC result (solid line) is somewhat different from the DSF one 
(dashed line) at $1.5 < q < 2.5$ fm$^{-1}$, 
although the two results are close to each other 
at backward angles $q > 2.5$ fm$^{-1}$; 
note that the dashed line agrees with the dot-dashed line 
at $1.5 < q < 2.5$ fm$^{-1}$. 
A few percent deviation between CDCC and DSF results in panel (b) 
for $\sigma_{\rm R}$ 
mainly comes from the difference 
between CDCC and DSF results 
in panel (a) at $1.5 < q < 2.5$ fm$^{-1}$. 
}
 \label{Fig-XSEC-He3Ni58-CDCC}
\end{center}
\end{figure}

Figure \ref{Fig-ReactionXSEC-He3Ni58-Pb208} shows $\sigma_{\rm R}$ 
for $^{3}$He scattering from $^{58}$Ni and $^{208}$Pb targets in a range 
of $E_{\rm in}/A_{\rm P}=30\mbox{--}150$~MeV.
The DSF model (circles) yields better agreement with the experimental data 
\cite{E32-56:Ingemarsson:2001,E31:Kadkin:1998} than the DF-FDA model (squares). 
For $^{58}$Ni target, the DSF model slightly underestimates the data 
around $E_{\rm in}/A_{\rm P}=30$~MeV, but this underestimation is 
solved by projectile breakup effects, as shown later in Fig. 
\ref{Fig-XSEC-He3Ni58-CDCC}(b).

Precisely speaking, the results of Fig. \ref{Fig-ReactionXSEC-He3Ni58-Pb208} 
depend on which $g$-matrix is taken. 
However, it is shown in Ref. \cite{Amos} that 
elastic scattering are mainly determined 
by the on-shell part of $g$-matrix. The Melbourne g-matrix is constructed 
from Bonn-B nuclear force \cite{Bonn-B}. We have confirmed that the g-matrix constructed 
from CD-Bonn nuclear force \cite{CD-Bonn} yields almost the same results as the Melbourne 
$g$-matrix; for example, the difference is only 5 mb 
for $^{3}$He+$^{208}$Pb scattering at $E_{\rm in}/A_{\rm P}=72.3$~MeV. 
We will make further discuss on this point in the forthcoming paper. 

\begin{figure}[htbp]
\begin{center}
 \includegraphics[width=0.40\textwidth,clip]{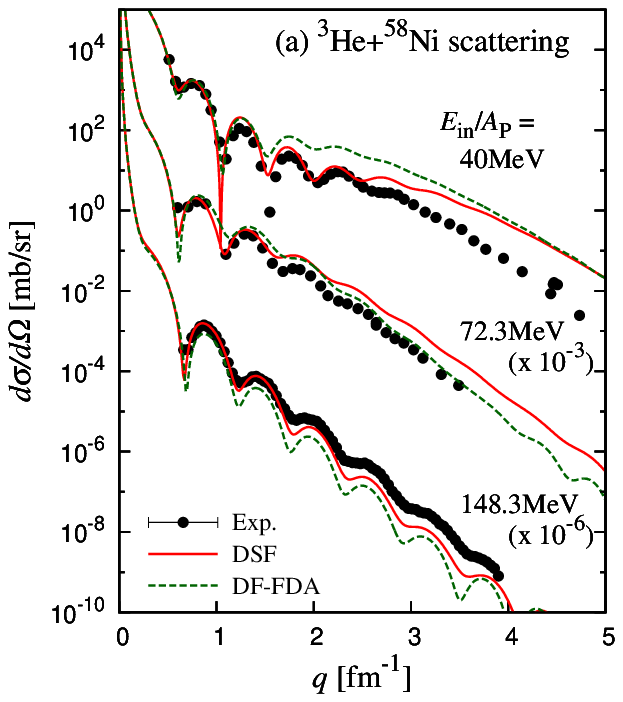}
 \includegraphics[width=0.40\textwidth,clip]{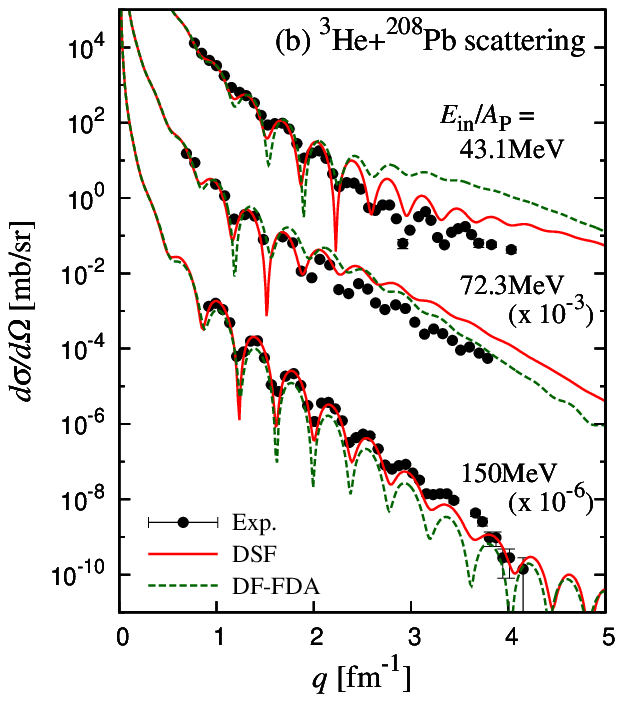}
 \caption{(Color online) 
Differential cross sections as a function of transfer momentum $q$ 
for (a) $^{3}$He+$^{58}$Ni scattering and 
(b) $^{3}$He+$^{208}$Pb scattering.
The cross section at each $E_{\rm in}/A_{\rm P}$ is multiplied by 
the factor shown in the panel. 
The solid and dashed lines denote results of the DSF and DF-FDA models, 
respectively. 
The spin-orbit force is not included in both the models. 
The experimental data are taken from Refs.~\cite{E40:Hyakutake:1980,E72:Willis:1973,E148:Kamiya:2003,E43:Djaloeis:1978,E150:Yamagata:1995}.
}
 \label{Fig-XSEC-He3Ni58-Pb208}
\end{center}
\end{figure}

\begin{figure}[htbp]
\begin{center}
 \includegraphics[width=0.40\textwidth,clip]{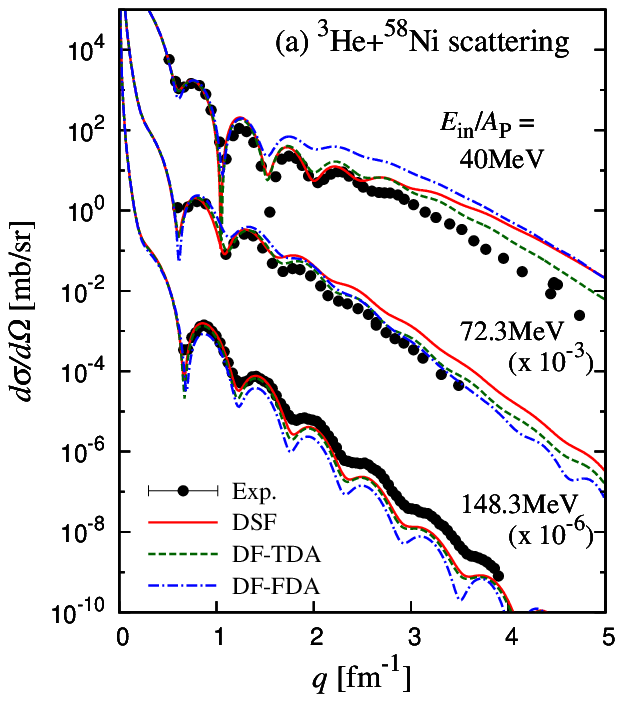}
 \includegraphics[width=0.40\textwidth,clip]{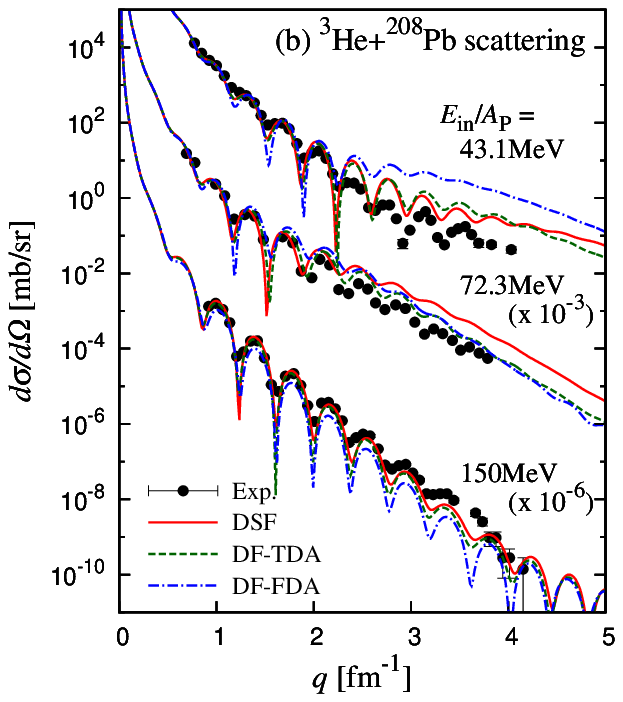}
 \caption{(Color online) 
Comparison of DSF, DF-TDA and DF-FDA models for 
differential cross sections in (a) $^{3}$He+$^{58}$Ni scattering and 
(b) $^{3}$He+$^{208}$Pb scattering.
The differential cross section at each $E_{\rm in}/A_{\rm P}$ is multiplied by 
the factor shown in the panel. 
The dashed (dot-dashed) lines denote results of the DF-TDA 
(DF-FDA) model, 
while the solid lines stand for results of the DSF model. 
The spin-orbit force is not included in both the models. 
The experimental data are taken from Refs.~\cite{E40:Hyakutake:1980,E72:Willis:1973,E148:Kamiya:2003,E43:Djaloeis:1978,E150:Yamagata:1995}.
}
 \label{Fig-XSEC-He3Ni58-Pb208-TDA}
\end{center}
\end{figure}

Figure \ref{Fig-XSEC-He3Ni58-Pb208} shows differential cross sections 
$d\sigma/d\Omega$ as a function of transfer momentum $q$ 
for $^{58}$Ni and $^{208}$Pb targets. 
For $E_{\rm in}/A_{\rm P} \approx 40$ and 150~MeV, the DSF model (solid line) 
definitely yields better agreement with the experimental data~\cite{E40:Hyakutake:1980,E72:Willis:1973,E148:Kamiya:2003,E43:Djaloeis:1978,E150:Yamagata:1995} than the DF-FDA model (dashed line). 
For $E_{\rm in}/A_{\rm P}=72$~MeV, agreement with the data is comparable 
between the two models; more precisely, the DSF model is better than 
the DF-FDA model at $q < 2$~fm$^{-1}$, 
but the latter is superior to the former at $q > 2$~fm$^{-1}$. 
The overestimation of the DSF model at $q > 2$~fm$^{-1}$ comes from 
the fact that knock-on processes are treated only approximately 
in the DSF model, as discussed later 
in Fig. \ref{Fig-XSEC-He3Ni58-Pb208-TDA}. 
The DSF model thus yield better description for $^{3}$He scattering than 
the DF-FDA model.

Figure \ref{Fig-XSEC-He3Ni58-CDCC} shows projectile-breakup and spin-orbit 
force effects on $d\sigma/d\Omega$ and $\sigma_{\rm R}$ 
for $^{3}$He+$^{58}$Ni scattering. 
In panel (a) for $d\sigma/d\Omega$, the dot-dashed and dashed 
lines stand for results of the DSF model with and without 
the spin-orbit force respectively, while 
the solid lines denote results of CDCC in which the spin-orbit 
force is neglected. Hence the difference between the dashed and solid lines 
shows projectile-breakup effects, while that  
between the dashed and dot-dashed lines corresponds to spin-orbit force effects. 
Both the effects are small, but improve agreement with the data. More precisely, spin-orbit force effects are 
appreciable at higher incident energies around 
$E_{\rm in}/A_{\rm P}=150$~MeV, but 
projectile-breakup effects are visible at lower incident energies such as 
$E_{\rm in}/A_{\rm P}=40$~MeV. The present result on spin-orbit force effects 
is consistent with the previous one of Ref. \cite{3He:Sakuragi}.

In panel (b) for $\sigma_{\rm R}$, 
the circles denote results of the DSF model 
without the spin-orbit force, while the triangles correspond to 
results of CDCC in which the spin-orbit force is neglected. 
In this panel, results of the DSF model with the spin-orbit force 
are not shown, since they agree 
with those of the DSF model without the spin-orbit force. 
This indicates that spin-orbit force effects are negligibly small. 
Projectile-breakup effects are shown by the difference between 
the triangle and the corresponding circle. The effects are more appreciable 
for $\sigma_{\rm R}$ than for $d\sigma/d\Omega$. The effects improve 
agreement with the experimental data 
particularly 
at low incident energies around $E_{\rm in}/A_{\rm P}=30$~MeV.

Comparing Fig. \ref{Fig-XSEC-He3Ni58-CDCC} with Figs. 
\ref{Fig-ReactionXSEC-He3Ni58-Pb208} and \ref{Fig-XSEC-He3Ni58-Pb208}, one 
can easily see that CDCC calculations based on the model Hamiltonian 
well describe $^{3}$He scattering and yield 
better agreement with the data than the DF-FDA model. 
The model Hamiltonian is thus good not only for deuteron and 
$^{4}$He scattering but also for $^{3}$He scattering.

\begin{figure}[htbp]
\begin{center}
 \includegraphics[width=0.40\textwidth,clip]{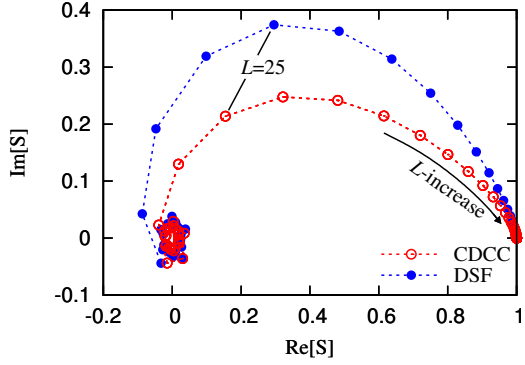}
 \caption{(Color online) 
Projectile-breakup effects on elastic $S$-matrix elements 
for $^{3}$He+$^{58}$Ni scattering at $E_{\rm in}/A_{\rm P}=40$~MeV. 
The filled (open) circles correspond the elastic $S$-matrix elements 
calculated with the DSF model (CDCC); here note that 
the spin-orbit force is not included. 
The elastic $S$-matrix elements tend to a point $(1,0)$ as 
$L$ increases.
}
 \label{Fig-Smat-He3Ni58}
\end{center}
\end{figure}

Figure \ref{Fig-Smat-He3Ni58} shows $L$ dependence of the elastic $S$-matrix 
elements $S_{L}$. The filled (open) circles denote the $S_{L}$ calculated 
with the DSF model (CDCC), where the spin-orbit force is neglected. 
Projectile-breakup effects shown by the difference between the open and 
filled circles are sizable for large $L$ but not for small $L$. 
At large $L$, furthermore, the effects reduce the absolute value of $S_{L}$. 
Therefore the dynamical polarization potential generated 
by $^{3}$He breakup is strongly absorptive in the peripheral 
region of T. This is the reason why $^{3}$He breakup effects are 
more appreciable for $\sigma_{\rm R}$ than for $d\sigma/d\Omega$.

In the DSF model, the nucleon-target potential $U_{i{\rm T}}$ is localized 
by the Brieva--Rook method. Now we discuss how 
the localization affects $^{3}$He scattering, although the localization 
is accurate for nucleon-target scattering itself as shown in Ref. 
\cite{Minomo:2009ds}. When the nucleon-target potential 
is not localized, the potential between P and T is obtained by 
\bea
U_{\rm DF}^{\rm TDA}=\langle  \Phi_0
| \sum_{i \in {\rm P}, j \in {\rm T}}g_{ij}({\rho_{\rm T}}) 
| \Phi_0 \rangle \; .
\label{Pot-DF-TDA}
\eea
This is nothing but the DF model with the TDA. This model is referred to 
as the DF-TDA model in this paper. 
The explicit form of DF-TDA potential is shown in Appendix 
\ref{Explicit forms of DF-FDA, DF-TDA and DSF potentials}, together 
with the explicit forms of DF-FDA and DSF potentials.

In the DF-TDA model, the nonlocality of the folding potential comes from 
knock-on exchange processes between interacting two nucleons, 
one in P and the other in T. 
As a consequence of the exchange, 
the one-body densities $\rho_{\rm P}(\vrr_{\rm P})$ and 
$\rho_{\rm T}(\vrr_{\rm T})$ are changed into the corresponding 
mixed densities ${\tilde \rho}_{\rm P}(\vrr_{\rm P},\vrr_{\rm P}-\vs)$ and 
${\tilde \rho}_{\rm T}(\vrr_{\rm T},\vrr_{\rm T}+\vs)$ as shown 
in Appendix 
\ref{Explicit forms of DF-FDA, DF-TDA and DSF potentials}, 
where 
$\vrr_{\rm P}$ ($\vrr_{\rm T}$) denotes the coordinate of nucleon 
in P (T) 
from the center of mass of P (T) and $\vs$ is the relative coordinate 
between interacting two nucleons. 
The one-body and mixed densities of P are defined by
\bea
\rho^{\mu}_{\rm P}(\vrr_{\rm P})&=&\sum_{a}
\phi^{* \mu}_{{\rm P}; a}(\vrr_{\rm P})
\phi^{\mu}_{{\rm P}; a}(\vrr_{\rm P}) ,\\
\tilde{\rho}^{\mu}_{\rm P}(\vrr_{\rm P},\vrr_{\rm P}-\vs)
&=&\sum_{a} 
\phi^{* \mu}_{{\rm P}; a}(\vrr_{\rm P})
\phi^{\mu}_{{\rm P}; a}(\vrr_{\rm P}-\vs) 
\eea
with the single particle wave function 
$\phi^{\mu}_{{\rm P}; a}(\vrr_{\rm P})$ of P characterized with 
the quantum number $a$, where $\mu$ denotes either proton or neutron. 
One can take the same definition also for the one-body and mixed densities 
of T.

The knock-on exchange is properly treated in the DF-TDA model. 
The resulting nonlocal potential between P and T can be localized 
with high accuracy by the Brieva--Rook method based on 
the local semi-classical approximation~\cite{Hag06}. 
In the DSF model, the change of $\rho_{\rm T}(\vrr_{\rm T})$ due to 
knock-on exchange is taken into account in the nucleon-target potential.  
However, the change of $\rho_{\rm P}(\vrr_{\rm P})$ is 
not treated as shown in Appendix 
\ref{Explicit forms of DF-FDA, DF-TDA and DSF potentials}, 
since the nucleon-target potential is localized. 
Therefore, one can see how the change of $\rho_{\rm P}(\vrr_{\rm P})$ due to
knock-on exchange affects $^{3}$He scattering, comparing results 
of the DSF model with those of the DF-TDA model.

Figure \ref{Fig-XSEC-He3Ni58-Pb208-TDA} shows differential cross 
sections calculated with the DSF, DF-TDA and DF-FDA models 
for (a) $^{3}$He+$^{58}$Ni scattering and 
(b) $^{3}$He+$^{208}$Pb scattering, 
where the spin-orbit force is neglected. 
The dashed (solid) lines correspond to results of the DF-TDA (DSF) model. 
The difference between the dashed and solid lines shows effects of 
the $\rho_{\rm P}$ change due to knock-on exchange. 
The effects are small for higher incident energies 
such as $E_{\rm in}/A_{\rm P}=150$~MeV. 
For lower incident energies less than $E_{\rm in}/A_{\rm P}=100$~MeV, 
meanwhile, the effects appear at larger angles and improve agreement with 
the experimental data, particularly 
in $q \ga 3$~fm$^{-1}$ for $^{3}$He+$^{58}$Ni scattering 
at $E_{\rm in}/A_{\rm P}=40$~MeV and in $q \ga 2$~fm$^{-1}$ 
for $^{3}$He+$^{58}$Ni scattering at $E_{\rm in}/A_{\rm P}=72$~MeV. 
The DF-TDA model thus yields better agreement with the data than 
the DSF model particularly at the larger angles. 
Eventually, the DF-TDA model (dashed line) well reproduce the data 
in a wide incident-energy range from $E_{\rm in}/A_{\rm P}=40$~MeV to 150~MeV, 
compared with the DF-FDA model (dot-dashed line). 
For $\sigma_{\rm R}$, furthermore, we have confirmed that the DF-TDA model 
yields almost the same result as the DSF model 
and hence better agreement with the data than DF-FDA model.

\begin{figure}[htbp]
\begin{center}
 \includegraphics[width=0.45\textwidth,clip]{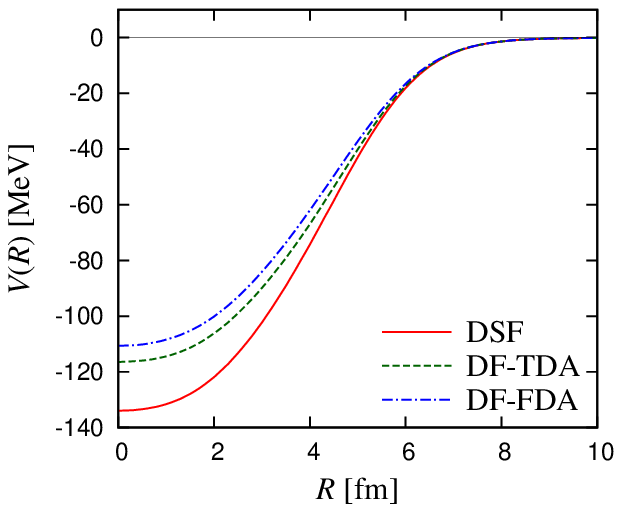}
 \includegraphics[width=0.45\textwidth,clip]{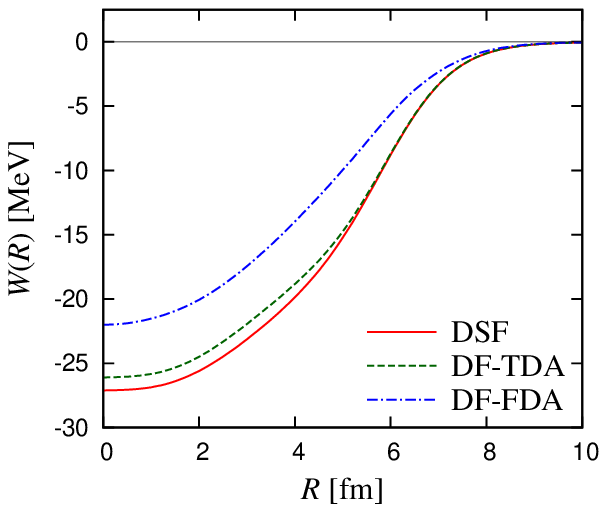}
 \caption{(Color online) 
Microscopic optical potentials for $^{3}$He+$^{58}$Ni scattering at 
$E_{\rm in}/A_{\rm P}=40$~MeV. Here $V(R)$ ($W(R)$) is 
the real (imaginary) part of the potentials. 
The dashed (dot-dashed) line stands 
for the DF-TDA (DF-FDA) potential, 
whereas the solid line denotes the DSF potential. 
}
 \label{Fig-U-He3Ni58E40}
\end{center}
\end{figure}

We compare DSF, DF-TDA and DF-FDA potentials 
in Fig. \ref{Fig-U-He3Ni58E40} for $^{3}$He+$^{58}$Ni scattering 
at $E_{\rm in}/A_{\rm P}=40$~MeV. 
The FDA has stronger Pauli-blocking effects than the TDA 
because of $\rho_{\rm P}+\rho_{\rm T} \ge \rho_{\rm T}$, 
so that the DF-FDA potential (dot-dashed line) is 
less attractive and less absorptive than the DF-TDA potential (dashed line). 
The DSF model (solid line) well simulates the DF-TDA potential 
in the peripheral region. This is the reason why 
the DSF model well simulates the DF-TDA model at forward angles of 
$q \la 3$~fm$^{-1}$.

Finally, we comment on a traditional Woods-Saxon (WS) parametrization 
for the phenomenological optical potential. 
For $^{3}$He scattering, it is difficult 
to reproduce the folding potential with a single WS form 
in the full range of $R$. In general, the forward scattering is sensitive to 
the peripheral region of the folding potential. If a single WS form 
is determined from the folding potential in the peripheral region, the 
single WS potential can reproduce the differential cross section at 
$q < 2$ fm$^{-1}$. If a double WS form is used, 
the potential is close to the folding potential and reproduces 
the differential cross section up to $q=4$ fm$^{-1}$. 
Thus the double WS form is recommendable for the present scattering. 

\section{Summary}
\label{Summary}

We have investigated how good the model Hamiltonian \eqref{model-H} is for 
$^{3}$He scattering. 
In the model Hamiltonian, the nucleon-target potential $U_{i{\rm T}}$ 
is obtained by 
folding the Melbourne $g$-matrix with the target density and localizing 
the resultant nonlocal folding potential with the Brieva--Rook method. 
In many calculations done so far, 
the phenomenological nucleon optical potential 
was used as $U_{i{\rm T}}$. As an advantage of the present approach 
from the previous one, the microscopic $U_{i{\rm T}}$ is obtainable even 
for the nucleon+target system in which no experimental data is available.    
As an advantage of the present approach from the DF-FDA model, 
the model Hamiltonian is consistent 
with the fact that the single Fermi sphere is considered 
when the $g$-matrix is evaluated in nuclear matter. 
In the present approach, projectile excitation effects 
have to be treated explicitly, 
but this can be done with CDCC since $U_{i{\rm T}}$ is localized.

The validity of the model Hamiltonian has been investigated 
in a wide incident-energy range of $E_{\rm in}/A_{\rm P}= 30\mbox{--}150$~MeV 
for heavier targets $^{58}$Ni and $^{208}$Pb, 
since the $g$-matrix is considered to be 
more reliable for heavier targets. 
CDCC calculations well reproduce the experimental data with no adjustable 
parameter. The calculations show that $^{3}$He breakup effects 
are appreciable only at lower incident energies around 40 MeV/nucleon and 
negligibly small 
at higher incident energies. Therefore the DSF model also well 
accounts for the experimental data. 
In fact, the DSF model yields better agreement with the data 
than the DF-FDA model. 
Thus we can propose that the DSF model is a practical model to describe 
$^{3}$He scattering and can conclude that the model Hamiltonian is 
good not only for deuteron and $^{4}$He scattering but also 
for $^{3}$He scattering. 

As a merit of the DSF model, the spin-orbit force between P and T can 
be easily evaluated. Effects of the force are found to be appreciable only 
at higher incident energies such as $E_{\rm in}/A_{\rm P}= 150$~MeV and 
negligibly small 
at lower incident energies. This is consistent with the previous 
analysis of Ref. \cite{3He:Sakuragi} based on the phenomenological 
nucleon-target potential.

When the nucleon-target potential is not localized, 
the DSF model becomes the DF-TDA model. 
Knock-on exchange processes are properly taken into account 
in the DF-TDA model, but the change of $\rho_{\rm P}(\vrr_{\rm P})$ due to 
knock-on exchange is not in the DSF model 
because the nucleon-target potential is localized. 
For lower incident energies such as  $E_{\rm in}/A_{\rm P} \la 70$~MeV, 
effects of the $\rho_{\rm P}$ change appear in differential cross sections 
at larger angles of $q \ga 2$~fm$^{-1}$ and improves agreement 
with the experimental data there. When one is interested 
in the backward angles, the DF-TDA model is better than DSF model. 
For higher incident energies such as $E_{\rm in}/A_{\rm P} = 150$~MeV, 
meanwhile, the effects are small and hence the DSF model is as good 
as the DF-TDA model.

In conclusion, the model Hamiltonian \eqref{model-H} works well 
for scattering of $0s$-shell nuclei such as deuteron, 
$^{3}$He and $^{4}$He. As a future work, it is quite interesting 
to investigate 
whether the model Hamiltonian is good also for scattering of 
$0p$-shell nuclei such as $^{12}$C and $^{16}$O.

\vspace*{5mm}

\section*{Acknowledgments}
We thank K.~Ogata and M.~Kohno for useful discussions. 
This work is supported in part by
by Grant-in-Aid for Scientific Research (No. 26400278)
from Japan Society for the Promotion of Science (JSPS). 

\vspace*{5mm}

\noindent
\appendix
\section{Explicit forms of DF-FDA, DF-TDA and DSF potentials}
\label{Explicit forms of DF-FDA, DF-TDA and DSF potentials}

Here we show the explicit forms of DF-FDA, DF-TDA and DSF potentials 
and discuss their relation. 
This is an overview of the previous discussion in Ref. \cite{Egashira:2014zda}.

The DF potential $U_{\rm DF}$ for P+T scattering at an incident energy 
$E_{\rm in}$ is composed of the direct and knock-on exchange 
terms, $U^{\rm DR}_{\rm DF}$ and $U^{\rm EX}_{\rm DF}$, defined 
by~\cite{DFM-standard-form,DFM-standard-form-2,Sum12} 
\bea
U^{\rm DR}_{\rm DF}(\vR) \hspace*{-0.15cm} &=& \hspace*{-0.15cm}
\sum_{\mu,\nu}\int \rho^{\mu}_{\rm P}(\vrr_{\rm P})
            \rho^{\nu}_{\rm T}(\vrr_{\rm T})
            g^{\rm DR}_{\mu\nu}(s;\rho_{\mu\nu}) d \vrr_{\rm P} d \vrr_{\rm T},
~~~~~
\label{eq:UD}
\\
U^{\rm EX}_{\rm DF}(\vR) \hspace*{-0.15cm} &=& \hspace*{-0.15cm}\sum_{\mu,\nu}
\int \tilde{\rho}^{\mu}_{\rm P}(\vrr_{\rm P},\vrr_{\rm P}-\vs)
\tilde{\rho}^{\nu}_{\rm T}(\vrr_{\rm T},\vrr_{\rm T}+\vs) \nonumber \\
            &&~~\hspace*{-1.0cm}\times g^{\rm EX}_{\mu\nu}(s;\rho_{\mu\nu}) \exp{[-i\vK(\vR) \cdot \vs/M]}
            d \vrr_{\rm P} d \vrr_{\rm T},~~~~
\label{eq:UEX}
\eea
where $\vs=\vrr_{\rm P}-\vrr_{\rm T}+\vR$ and $\mu$ and $\nu$ denote 
either proton or neutron. 
The non-local $U^{\rm EX}$ has been localized 
with the local semi-classical approximation~\cite{Brieva-Rook}, 
where the local momentum $\hbar \vK(\vR)$ between P and T is 
given by $\hbar K(R) \equiv \sqrt{2 M (E_{\rm in} -U_{\rm DF}(R))}$ 
with the reduced mass $M=A_{\rm P} A_{\rm T}/(A_{\rm P} +A_{\rm T})$ of 
the P+T system. The direct and exchange parts, $g^{\rm DR}_{\mu\nu}$ and
$g^{\rm EX}_{\mu\nu}$, of $g$-matrix depend on the density 
$\rho_{\mu\nu}$ defined by  
\bea
 \rho_{\mu\nu}=\rho^{\mu}_{\rm P}(\vrr_{\rm P}-\vs/2)
 +\rho^{\nu}_{\rm T}(\vrr_{\rm T}+\vs/2)
\label{local-density approximation-1}
\eea
in the FDA and 
\bea
 \rho_{\mu\nu}=\rho^{\nu}_{\rm T}(\vrr_{\rm T}+\vs/2)
\label{local-density approximation-2}
\eea
in the TDA. 
The difference between the DF-FDA and DF-TDA stems from only 
that between 
Eqs. \eqref{local-density approximation-1} and 
\eqref{local-density approximation-2}. 

For N+T scattering at an incident energy $E_{\rm in}^{\rm N}$, 
the SF potentials $U_{\mu}$ 
for proton ($\mu=-1/2$) and neutron ($\mu=1/2$) scattering also 
consist of the direct and knock-on exchange terms, 
$U_{\mu}^{\rm DR}$ and $U_{\mu}^{\rm EX}$, defined by~\cite{Minomo:2009ds}
\bea
\label{eq:UD-NA}
U_{\mu}^{\rm DR}(\vrr_{\mu}) \hspace*{-0.15cm} &=& \hspace*{-0.15cm}
\sum_{\nu}\int 
            \rho^{\nu}_{\rm T}(\vrr_{\rm T})
            g^{\rm DR}_{\mu\nu}(s;\rho_{\mu\nu}) d \vrr_{\rm T}, 
\\ \label{eq:UEX-NA}
U_{\mu}^{\rm EX}(\vrr_{\mu}) \hspace*{-0.15cm} &=& \hspace*{-0.15cm}\sum_{\nu}
\int 
\tilde{\rho}^{\nu}_{\rm T}(\vrr_{\rm T},\vrr_{\rm T}+\vs) \nonumber \\
            &&~~\hspace*{-1.0cm}\times g^{\rm EX}_{\mu\nu}(s;\rho_{\mu\nu}) 
\exp{[-i\vK_{\mu}(\vrr_{\mu}) \cdot \vs]}
            d \vrr_{\rm T}~~~~
\eea
for $\vs=\vrr_{\mu}-\vrr_{\rm T}$, where $\vrr_{\mu}$ is the coordinate 
of an incident nucleon N from the center of mass of T and  
the local momentum $\hbar \vK_{\mu}(\vrr_{\mu})$ between N and T 
is given by 
$\hbar K_{\mu}(r_{\mu}) \equiv \sqrt{2\mu_{\rm NT} 
(E_{\rm in}^{\rm N} -U_{\mu}(r_{\mu}))}$ for 
the reduced mass $\mu_{\rm NT}$ of the N+T system.

The DSF potential $U_{\rm DSF}$ is obtained by folding 
the nucleon-target potentials $U_{\mu}^{\rm DR}+U_{\mu}^{\rm EX}$ with 
the projectile density $\rho^{\mu}_{\rm P}$. The direct and knock-on 
exchange parts are then given by 
\bea
U^{\rm DR}_{\rm DSF}(\vR) \hspace*{-0.15cm} &=& \hspace*{-0.15cm}
\sum_{\mu} \int \rho^{\mu}_{\rm P}(\vrr_{\rm P}) 
U_{\mu}^{\rm DR}(\vR+\vrr_{\rm P})
             d \vrr_{\rm P} 
\nonumber \\  
&=& \hspace*{-0.15cm}
\sum_{\mu,\nu}\int \rho^{\mu}_{\rm P}(\vrr_{\rm P})
            \rho^{\nu}_{\rm T}(\vrr_{\rm T})
           g^{\rm DR}_{\mu\nu}(s;\rho_{\mu\nu}) d \vrr_{\rm P} d \vrr_{\rm T} 
\label{eq:DSF-UD-2},
\\ 
U^{\rm EX}_{\rm DSF}(\vR) \hspace*{-0.15cm} &=& \hspace*{-0.15cm}\sum_{\mu}
\int \rho^{\mu}_{\rm P}(\vrr_{\rm P}) 
U_{\mu}^{\rm EX}(\vR+\vrr_{\rm P}) d \vrr_{\rm P}
\nonumber \\ 
 \hspace*{-0.15cm} &=& \hspace*{-0.15cm}\sum_{\mu,\nu}
\int {\rho}^{\mu}_{\rm P}(\vrr_{\rm P})
\tilde{\rho}^{\nu}_{\rm T}(\vrr_{\rm T},\vrr_{\rm T}+\vs) \nonumber \\
            &&~~\hspace*{-1.0cm}\times g^{\rm EX}_{\mu\nu}(s;\rho_{\mu\nu}) \exp{[-i\vK_{\mu}(\vrr_{\mu}) \cdot \vs/M]}
            d \vrr_{\rm P} d \vrr_{\rm T},~~~~~~
\label{eq:DSF-UEX-2}
\eea

Now we consider heavy targets satisfying 
$A_{\rm T}\gg A_{\rm P}>1$ for simplicity. 
The $g$ matrix depends on an energy of nucleon in P. 
The energy is $E_{\rm in}^{\rm N}$ for N+T scattering and 
$E_{\rm in}/A_{\rm P}$ for P+T scattering. We can then find 
that $U^{\rm DR}_{\rm DF-TDA}=U^{\rm DR}_{\rm DSF}$ 
when $E_{\rm in}^{\rm N}=E_{\rm in}/A_{\rm P}$. 
In the peripheral region of T important for elastic scattering, 
the local momenta $\hbar \vK_{\mu}(\vrr_{\mu})$ and $\hbar \vK(\vR)$ are close 
to their asymptotic values, $\hbar \vK_{\mu}(\infty)$ and $\hbar \vK(\infty)$, 
respectively. When $E_{\rm in}^{\rm N}=E_{\rm in}/A_{\rm P}$, 
the asymptotic values satisfy 
\bea
\vK_{\mu}(\infty)=\vK(\infty)/M .
\label{K-relation}
\eea
Therefore we see that $U^{\rm DR}_{\rm DF-TDA}=U^{\rm DR}_{\rm DSF}$ 
if $\tilde{\rho}^{\mu}_{\rm P}(\vrr_{\rm P},\vrr_{\rm P}-\vs)$ 
is identical with ${\rho}^{\mu}_{\rm P}(\vrr_{\rm P})$. 
The approximation 
$\tilde{\rho}^{\mu}_{\rm P}(\vrr_{\rm P},\vrr_{\rm P}-\vs) 
\approx \tilde{\rho}^{\mu}_{\rm P}(\vrr_{\rm P},\vrr_{\rm P}) = 
\rho^{\mu}_{\rm P}(\vrr_{\rm P})$ is known to be good 
in the peripheral region of T~\cite{Minomo:2009ds}.

\vspace*{10cm}



\begin{thebibliography}{00}


\bibitem{CDCC-review1}
M.~Kamimura, 
M.~Yahiro, Y.~Iseri, Y.~Sakuragi, H.~Kameyama, and
M.~Kawai, \newblock
Prog.\ Theor.\ Phys.\ Suppl.\ {\bf 89}, 1 (1986).

\bibitem{CDCC-review2}
N.~Austern, 
Y.~Iseri, M.~Kamimura, M.~Kawai, G.~Rawitscher, and
M.~Yahiro, \newblock
Phys.\ Rep.\ {\bf 154}, 125 (1987).

\bibitem{CDCC-review3} 
  M.~Yahiro, K.~Ogata, T.~Matsumoto, and K.~Minomo,
 Prog. Theor. Exp. Phys. {\bf 2012}, 01A206 (2012).


\bibitem{M3Y}
G.~Bertsch, J.~Borysowicz, H.~McManus, and W.~G.~Love,
Nucl. Phys. A {\bf 284}, 399 (1977).

\bibitem{JLM}
J.~-P.~Jeukenne, A.~Lejeune, and C.~Mahaux, Phys. Rev. C{\bf 16}, 80 (1977);\\
J.~-P.~Jeukenne, A.~Lejeune, and C.~Mahaux, Phys. Rep. {\bf 25}, 83 (1976).


\bibitem{Brieva-Rook}
F.~A.~Brieva and J.~R.~Rook, Nucl. Phys. A {\bf 291}, 299 (1977);
{\it ibid.} 291, 317 (1977); {\it ibid.} 297, 206 (1978).

\bibitem{Satchler-1979}
G.~R.~Satchler and W.~G.~Love, Phys. Rep. {\bf 55}, 183 (1979).

\bibitem{Satchler}
G.~R.~Satchler, \lq\lq Direct Nuclear Reactions'',
Oxfrod University Press, (1983).

\bibitem{CEG}
N.~Yamaguchi, S. ~Nagata, and T.~Matsuda, Prog. Theor.
Phys. {\bf 70}, 459 (1983);
N.~Yamaguchi, S.~Nagata, and J.~Michiyama,
Prog. Theor. Phys. {\bf 76}, 1289 (1986).

\bibitem{Rikus-von Geramb}
L.~Rikus, K.~Nakano, and H.~V.~Von Geramb, Nucl. Phys. A {\bf 414}, 413 (1984);
L.~Rikus, and H.~V.~Von Geramb, Nucl. Phys. A {\bf 426}, 496 (1984).

\bibitem{Amos}
K.~Amos, P.~J.~Dortmans, H.~V.~Von Geramb, S.~Karataglidis, and J.~Raynal, in \textit{Advances in Nuclear Physics}, edited by
J.~W.~Negele and E.~Vogt(Plenum, New York, 2000) Vol. 25, p. 275.

\bibitem{CEG07}
T.~Furumoto, Y.~Sakuragi, and Y.~Yamamoto, Phys. Rev. C {\bf 78}, 044610 (2008).

\bibitem{MP}
Y.~Yamamoto, T.~Furumoto, N.~Yasutake, and Th.~A.~Rijken,
Phys. Rev. C {\bf 88}, 022801 (2013).

\bibitem{Saliem}
S.~M.~Saliem and W.~Haider, J. Phys. G {\bf 28}, 1313 (2002).

\bibitem{DFM-standard-form}
B. Sinha, Phys. Rep. {\bf 20}, 1 (1975). \\
B. Sinha and S. A. Moszkowski, Phys. Lett. B{\bf 81}, 289 (1979).


\bibitem{Arellano:1995}
H.~F.~Arellano, F.~A.~Brieva, and W.~G.~Love, 
Phys. Rev. C {\bf 52}, 301 (1995). 

\bibitem{rainbow}
 D.~T.~Khoa, 
 W.~von Oertzen, H.~G.~Bohlen, and S.~Ohkubo,
 J. Phys. G {\bf 34}, R111 (2007).

\bibitem{DFM-standard-form-2}
T. Furumoto, Y. Sakuragi, and Y. Yamamoto, Phys. Rev. C{\bf 82}, 044612 (2010).


\bibitem{Sum12}
T.~Sumi {\it et al.}, 
Phys. Rev. C {\bf 85}, 064613 (2012). 



\bibitem{Minomo:2009ds}
  K.~Minomo, K.~Ogata, M.~Kohno, Y.~R.~Shimizu, and M.~Yahiro,
  J.\ Phys.\ G {\bf 37}, 085011 (2010)
  [arXiv:0911.1184 [nucl-th]].

\bibitem{Hag06}
K. Hagino, T. Takehi, and N. Takigawa,
Phys. Rev. C {\bf 74} (2006), 037601.



\bibitem{Koning-Delaroche}
A.~J.~Koning and J.~P.~Delaroche, Nucl. Phys. A {\bf 713} 231 (2003).



\bibitem{Dirac1}
S.~Hama, B.~C.~Clark, E.~D.~Cooper, H.~S.~Sherif, and R.~L.~Mercer, Phys. Rev. C {\bf 41}, 2737 (1990).

\bibitem{Dirac2}
E.~D.~Cooper, S.~Hama, B.~C.~Clark, and R.~L.~Mercer, Phys. Rev. C {\bf 47}, 297 (1993).


\bibitem{Perey-Perey}
C.~M.~Perey and F.~G.~Perey, At. Data Nucl. Data Tables {\bf 17}, 1 (1976).



\bibitem{Toyokawa:2013uua}
  M.~Toyokawa, K.~Minomo, and M.~Yahiro,
 Phys. Rev. C{\bf 88}, 054602 (2013).




\bibitem{Yahiro-Glauber}
M.~Yahiro, K.~Minomo, K.~Ogata, and M.~Kawai,
Prog. Theor. Phys. {\bf 120}, 767 (2008).



\bibitem{Watson}
K.~M.~Watson, Phys. Rev. {\bf 89}, 575 (1953).

\bibitem{KMT}
A.~K.~Kerman, H.~McManus, and R.~M.~Thaler, Ann.
Phys. {\bf 8}, 551 (1959).

\bibitem{Izu80}
T. Izumoto, S. Krewald, and A. Faessler,
Nucl. Phys. A {\bf 341}, 319 (1980).



\bibitem{Watanabe-Mg-2014} 
S.~Watanabe {\it et al.}, Phys. Rev. C {\bf 89}, 044610 (2014). 


\bibitem{Takechi-2010}
M. Takechi {\it et al.}, Phys. Lett. B {\bf 707}, 357 (2010).

\bibitem{Takechi-Mg}
M. Takechi {\it et al.}, Phys. Rev. C {\bf 90}, 061305(R) (2014).




\bibitem{Min11}
K.~Minomo, T.~Sumi, M.~Kimura, K.~Ogata, Y.~R.~Shimizu, and M.~Yahiro,
Phys. Rev. C {\bf 84}, 034602 (2011).

\bibitem{Min12}
K.~Minomo, T.~Sumi, M.~Kimura, K.~Ogata, Y.~R.~Shimizu, and M.~Yahiro,
Phys. Rev. Lett. {\bf 108}, 052503 (2012).



\bibitem{Egashira:2014zda} 
  K.~Egashira, K.~Minomo, M.~Toyokawa, T.~Matsumoto and M.~Yahiro,
  Phys.\ Rev.\ C {\bf 89}, 064611 (2014). 


\bibitem{Takeda}
G. Takeda and K. M. Watson, Phys. Rev. {\bf 97}, 1336(1955).

\bibitem{Picklesimer}
A. Picklesimer and R. M. Thaler, Phys. Rev. C{\bf 23}, 42(1981).


\bibitem{3He:Petrovich}
F.~Petrovich, R.~J.~Philpott, A.~W.~Carpenter and J.~A.~Carr, 
Nucl. Phys. A {\bf 425}, 609 (1984).

\bibitem{3He:Cook}
J.~Cook, Nucl. Phys. A {\bf 465}, 207 (1987).

\bibitem{3He:Sakuragi}
Y.~Sakuragi, M.~Katuma, Nucl. Instrum. Methods Phys. Res. A {\bf 402}, 
347 (1998).

\bibitem{Minnesota-force}
Y. C. Tang, M. LeMere, and D. R. Thompson, Phys. Rep. {\bf 47},
167 (1978).

\bibitem{C12-density}
H.~de Vries, C.~W.~de Jager, and C.~de Vries,
At. Data Nucl. Data Tables {\bf 36}, 495 (1987).



\bibitem{Singhal}
R.~P.~Singhal, M.~W.~S.~Macauley, and P.~K.~A.~De Witt Huberts, Nucl. Instr. and Meth. {\bf 148}, 113 (1978).

\bibitem{GognyD1S}
J.~F.~Berger, M.~Girod, and D.~Gogny,
Comput. Phys. Commun. {\bf 63}, 365 (1991). 


\bibitem{CC:Khoa}
D.~T.~Khoa and D.~C.~Cuong, Phys. Lett. B {\bf 660}, 331 (2008).

\bibitem{CC:Furumoto}
T.~Furumoto and Y.~Sakuragi, Phys. Rev. C {\bf 87}, 014618 (2013).



\bibitem{E32-56:Ingemarsson:2001}
A.~Ingemarsson \textit{et al}., Nucl. Phys. A {\bf 696}, 3 (2001).

\bibitem{E31:Kadkin:1998}
E.~P.~Kadkin \textit{et al}., Phys. Atom. Nucl. {\bf 61}, 1459 (1998).



\bibitem{Bonn-B}
R.~Machleidt, K.~Holinde, and Ch.~Elster, Phys. Rep. {\bf 149}, 1 (1987).

\bibitem{CD-Bonn}
R.~Machleidt, Phys. Rev. C {\bf 63}, 024001 (2001).



\bibitem{E40:Hyakutake:1980}
M.~Hyakutake \textit{et al}., Nucl. Phys. A {\bf 333}, 1 (1980).

\bibitem{E72:Willis:1973}
N.~Willis, I.~Brissaud, Y.~Le~Bornec, B.~Tatischeff, and G.~Duhamel, Nucl. Phys. A {\bf 204}, 454 (1973).

\bibitem{E148:Kamiya:2003}
J.~Kamiya {\textit et al}., Phys. Rev. C {\bf 67}, 064612 (2003).

\bibitem{E43:Djaloeis:1978}
A.~Djaloeis, J.~-P.~Didelez, A.~Galonsky, and W.~Oelert, Nucl. Phys. A {\bf 306}, 221 (1978).

\bibitem{E150:Yamagata:1995}
T.~Yamagata {\textit et al}., Nucl. Phys. A {\bf 589}, 425 (1995).





\end{thebibliography}
\end{document}